# Interpretation of TOVS Water Vapor Radiances Using a Random Strong Line Model

Brian J. Soden
*Geophysical Fluid Dynamics Laboratory*
*National Oceanic and Atmospheric Administration*

Francis P. Bretherton
*Space Science and Engineering Center*
*University of Wisconsin-Madison*

**Abstract**

This study illustrates the application of a random strong line (RSL) model of radiative transfer to the interpretation of satellite observations of the upwelling radiation in the 6.3 µm water vapor absorption band. The model, based upon an assemblage of randomly overlapped, strongly absorbing, pressure broadened lines, is compared to detailed radiative transfer calculations of the upper (6.7 µm) tropospheric water vapor radiance and demonstrated to be accurate to within ~ 1.2 K. Similar levels of accuracy are found when the model is compared to detailed calculations of the middle (7.3 µm) and lower (8.3 µm) tropospheric water vapor radiance, provided that the emission from the underlying surface is taken into account. Based upon these results, the RSL model is used to interpret TOVS-observed water vapor radiances in terms of the relative humidity averaged over deep layers of the upper, middle, and lower troposphere. We then present near-global maps of the geographic distribution and climatological variations of upper, middle and lower tropospheric humidity from TOVS for the period 1981-1991. These maps clearly depict the role of large-scale circulation in regulating the location and temporal variation of tropospheric water vapor.

# 1. Introduction

Water vapor plays a fundamental role in regulating the earth's climate both directly through its effect on the radiation balance and indirectly through its coupling with other components of the hydrologic cycle. Yet despite its importance, significant gaps in our knowledge of atmospheric moisture still exist. One source of information on tropospheric water vapor which has been largely overlooked for climate studies are water vapor radiances provided by the TIROS Operational Vertical Sounder (TOVS). The TOVS instrument package provides measurements of the upwelling radiance in a range spectral channels, three of which are located within the 6.3 µm water vapor absorption band. These channels, located at 6.7 µm, 7.3 µm, 8.3 µm, are sensitive to the amount of water vapor integrated over broad layers centered in the upper (200-500 mb), middle (400-800 mb), and lower troposphere (700mb - surface), respectively (see figure 1). Two aspects of the TOVS water vapor radiances make them well suited for climatological studies. One is their high spatial and temporal coverage. The TOVS instrument is carried onboard the NOAA series of polar orbiting satellites each of which provide near global coverage every 12 hours. Since there are typically two satellites in operation at any given time, measurements of the TOVS water vapor radiances are usually available 2-4 times per day for most regions. The second advantageous feature of TOVS radiances is the availability of a lengthy data set. A temporally coherent archive of TOVS water vapor radiances exists from February 1979 to present. The current 15+ year archive covers 3 El Nino episodes and provides an invaluable time series for describing the distribution and variation of moisture on time scales from days to years. This combination of high spatial/temporal coverage and lengthy duration is unmatched by any other water vapor archive.

Despite these advantages there are nevertheless some limitations to the TOVS water vapor radiances. Since the measurements are sensitive to the integrated amount of moisture over layers several kilometers thick, TOVS radiances provide only limited information on the vertical distribution of moisture. While this has long been a problem for assimilating the information into numerical models, it is less of a hindrance for climatological studies which are often interested in the characteristics of moisture averaged over a finite depth of the atmosphere rather than at a single level. Another limitation is that, due to the strong attenuation of infrared radiation by clouds, most information on water vapor is negated when clouds obscure the satellite pixel of interest. The impact of this problem is ameliorated by the relatively high horizontal resolution of the TOVS water vapor channels (~20 km at nadir) which enables measurements to be obtained in conditions of up to 75% cloud cover. Still this effect is likely to introduce a clear sky bias into the data set, the magnitude of which is difficult to estimate. Preliminary analysis of the difference between TOVS and radiosonde humidity climatologies as a function of cloud cover suggests that this sampling limitation introduces a slight dry bias in the TOVS water vapor measurements of $< 10\%$, expressed in terms of the relative humidity, with no distinct geographic dependence (Soden et al., 1995).

Despite the availability of extensive archives of TOVS water vapor radiances, they have been largely overlooked for climate studies with a few recent exceptions (Wu et al., 1993; Salathe et al., 1994; Soden and Fu, 1995; Bates and Wu, 1995). One reason for the lack of attention is due to the difficulty of interpreting the observed radiances in terms of a more familiar water vapor quantity. In their study of the GOES 6.7 μm channel, Soden and Bretherton (1993) (hereafter referred to as SB93) addressed this issue by developing an interpretation tool based upon a theoretical model of radiative transfer in the 6.3 μm water vapor absorption band. Rather than attempting retrievals in the form of vertical soundings, attention was focused on the information actually present in the observations, minimizing extraneous assumptions. This random strong line (RSL) model, which is summarized in the next section, provides a convenient method for interpreting 6.7 μm radiances in terms of the vertically averaged upper tropospheric relative humidity.

The purposes of this paper are:

(i) to apply the RSL model to TOVS upper tropospheric water vapor radiances (6.7 μm).

(ii) to extend application of the RSL model to middle (7.3 μm) and lower (8.3 μm) tropospheric water vapor radiances, including an allowance for the surface emission.

(iii) to present near-global maps of the upper, middle, and lower tropospheric relative humidity inferred from TOVS and explore their seasonal and interannual variations.

(iv) to examine the sensitivity of the inferred moisture distribution to the empirical tuning of the RSL model.

The remaining portion of this paper is organized as follows: Section 2 briefly reviews the RSL model developed by SB93. Section 3 discusses the application and verification of the RSL model for the TOVS upper, middle, and lower tropospheric water vapor radiances. Section 4 presents maps of the climatological variations of upper, middle, and lower tropospheric humidity inferred from TOVS. Section 5 describes a sensitivity analysis of the RSL model and the conclusions of this study are summarized in section 6.

## 2. The Random Strong Line Model

This section summarizes aspects of the random strong line model relevant to the present study - the reader is referred to SB93 for complete details. To facilitate the interpretation of satellite-observed water vapor radiances, SB93 developed a simplified model of radiative transfer based upon a set of strongly-absorbing, pressure-broadened lines. Following Goody [1964] each narrow, irregularly-spaced line is assumed to behave as an independent isolated absorber and the assemblage is then modeled as a large number of randomly located, overlapping lines. Using this random strong line (RSL) model and considering an idealized atmospheric profile in which the relative humidity $r$ and dimensionless lapse rate $\beta = p/T\, dT/dp$ are constant over the range of pressures to which the water vapor channel is sensitive, a simple analytical expression for the brightness temperature ($T_b$) may be obtained (see eq. 20 of SB93)

$$a + b \cdot T_b = \log_e\left(\frac{r p_o}{\beta \cos\theta}\right) \qquad (1)$$

The variables comprising the right hand side of equation (1) are the relative humidity $r$, lapse rate

$\beta$, satellite zenith angle $\theta$, and a reference pressure $p_o$ which is equal to the pressure of the 240 K isotherm divided by 300 mb. This expression, which is the central result of the RSL model proposed by SB93, provides a simple means of interpreting the brightness temperature in terms of more familiar atmospheric quantities. It demonstrates that for a profile with constant relative humidity and lapse rate, the brightness temperature varies logarithmically with the ratio ($rp_o/\beta\cos\theta$). Changes in relative humidity represent the largest source of variation in brightness temperature with a factor of 2 increase in $r$ resulting in a decrease in $T_b$ of $\log(2)/b \cong 6$ K. The normalized base pressure $p_o$ varies between the tropics ($p_o \cong 0.9$) and mid-latitudes ($p_o \cong 1.5$) by +/- 25% and thus also constitutes an important source of variation in $T_b$. Fortunately $p_o$ varies in a quite predictable manner. Figure 2a shows the zonally average distribution of $p_o$ for January and July as determined from ECMWF analyses. This figure illustrates that the dominant sources of variation in $p_o$ reflect the seasonal and latitudinal shifts in the mean temperature structure of the troposphere. These variations are easily determined from temperature climatologies and thus their impact upon the $T_b$ can be readily accounted for. Variations in atmospheric lapse rate $\beta \cong 0.21$ +/- .01 are typically a factor of two or more smaller than variations in $p_o$ and exhibit a less systematic geographic distribution (figure 2b), consequently they are of less importance for interpreting the water vapor radiances. Thus, to a reasonable degree of accuracy (+/- 0.5 K) variations in $\beta$ may be neglected and a modified form of equation (1) can be used,

$$a + b \cdot T_b = \log_e\left(\frac{rp_o}{\cos\theta}\right) \qquad (2)$$

This simplified form of the RSL model predicts that the water vapor $T_b$ measures essentially the logarithm of the mean relative humidity averaged over broad layers of the troposphere subject to a straight-forward allowance for viewing angle ($\cos\theta$) and a minor correction for the mean temperature structure ($p_o$).

This form of the RSL model differs slightly from that used by SB93 for the GOES 6.7 μm channel (see equation 23 of SB93). Namely, SB93 did not include variations in $p_o$ in their final relationship (although its importance was expected), noting that it did not significantly improve the performance of the RSL model. The reason for this was not well understood by SB93 but now is believed to be a spurious result related to the GOES viewing geometry. Since the GOES platform is geostationary, higher latitudes are viewed with a systematically larger zenith angle, particularly for mid to high latitudes. As a result, the layer which the GOES $T_{6.7}$ senses shifts towards higher levels of the upper troposphere. Since relative humidity within the upper troposphere typically decreases with height, high zenith angle measurements (e.g. GOES) sense lower relative humidities compared to low zenith angle measurements (e.g. TOVS) and this effect acts to compensate for the increase in $p_o$ with latitude. Thus for GOES, the expected variations in $p_o$ were masked by the systematic variations in viewing angle with latitude combined with a decrease in relative humidity with height in the upper troposphere.

In deriving equation (2) it is also assumed that the atmosphere is sufficiently opaque such that the surface contribution to the observed $T_b$ is negligible. While this assumption is generally valid for the upper tropospheric water vapor channel (6.7 µm) for which the RSL model was originally derived, it is not always true for the middle (7.3 µm) and lower (8.3 µm) tropospheric water vapor channels. In situations where the surface emission is an important component of the observed $T_b$, it is necessary to adjust the $T_b$ to account for the surface contribution. This surface adjustment is analogous to the procedure described by SB93 to account for the radiance contribution of underlying clouds in the 6.7 µm channel (see section 4.3 of SB93 for details). Figure 3 shows the brightness temperature correction (ΔT) as a function of the difference between the surface temperature ($T_s$) and the brightness temperature $T_b$. When the $T_b$ is at least 20 K cooler than $T_s$, the surface contribution is negligible and no correction is necessary. For the latitudes of interest in this study (60N-60S), this is virtually always the case for the upper tropospheric channel and is the norm for the middle tropospheric channel. The lower tropospheric channel, on the other hand, is rarely 20 K cooler than $T_s$ and hence a correction is almost always necessary. At the other extreme, when the difference between the $T_s$ and $T_b$ is only 7 K, the surface correction is also 7 K.

That is, the correction ΔT is now as large as the signal from the water vapor above the surface and this is roughly the upper limit for which a $T_b$ correction is reasonable.

## 3. Verification of the RSL Model

In this section we investigate the validity of the RSL model for interpreting TOVS water vapor radiances by comparing the simple logarithmic relationship predicted by the RSL model with detailed radiative transfer calculations using known profiles of temperature and moisture.

### a. Upper Tropospheric Water Vapor Radiances (6.7 μm)

The 6.7 μm channel is located near the center of the 6.3 μm water vapor absorption band and under clear sky conditions is sensitive primarily to the relative humidity averaged over a broad layer centered in the upper troposphere (roughly 200-500 mb). Figure 1 illustrates the sensitivity of the 6.7 μm channel to changes in relative humidity at various levels for a typical tropical profile viewed at zenith.

The RSL model predicts that the brightness temperature at 6.7 μm ($T_{6.7}$) should vary logarithmically with the ratio $\bar{r}_u p_o / \cos\theta$, where $\bar{r}_u$ represents the vertically averaged relative humidity in the upper troposphere weighted according to the 6.7 μm sensitivity curve (e.g. figure 1) and sum to unity. The weighting profile varies slightly depending upon the precise nature of the temperature and moisture profile, being slightly higher for warm/moist profiles and slightly lower for cold/dry profiles. Table 1 lists weights for computing $\bar{r}_u$ from the relative humidity at standard pressure levels. The weights are categorized according to the pressure $p_{6.7}$ corresponding to the channel brightness temperature $T_{6.7}$ to allow for shifts in the layer to which the $T_{6.7}$ is sensitive. To evaluate the validity of the RSL approximation, we compare the relationship predicted in equation (2) with detailed calculations of the 6.7 μm radiance using the CIMSS (Cooperative Institute for Meteorological Satellite Studies) transmittance model. The CIMSS transmittance code is a 40-level multivariate regression model based upon FASCODE3 line-by-line calculations and is functionally similar to that described by Eyre (1991). Figure 4a compares synthetic $T_{6.7}$ calculated from the CIMSS transmittance model for ~8,000 different atmospheric profiles versus the corresponding value of $\log(\bar{r}_u p_o/\cos\theta)$, where $\bar{r}_u$ is calculated explicitly from the same set of profiles and $p_o$ is obtained from climatology. To provide a sample representative of a wide range of atmospheric conditions, the temperature and moisture profiles are obtained from ECMWF operational analyses for July 1, 1988 using all profiles equatorward of $60^o$ latitude. A zenith viewing geometry ($\theta=0$) is chosen for all calculations to be consistent with the TOVS limb-corrected radiances[1] examined in section 4.

The radiance calculations reveal a very strong correlation (r=-0.96) between the forward calculations of $T_{6.7}$ and the corresponding value of $\log(\bar{r}_u p_o/\cos\theta)$. Thus the simple relationship predicted by the RSL model explains greater than 90% of the variability in $T_{6.7}$ in terms of the vertically averaged upper tropospheric relative humidity $\bar{r}_u$. The correlation is negative indicating that as the relative humidity increases, the $T_{6.7}$ decreases. The scatter in $T_{6.7}$ for a given $\bar{r}_u$ is +/- 1.2 K which corresponds to an uncertainty in $\bar{r}_u$ of ~10%. The good agreement of the radiance calculations with the logarithmic relation predicted by the RSL model supports the validity of the simple relationship and demonstrates that, to a reasonable degree of accuracy, measurements of $T_{6.7}$ can be interpreted in terms of the relative humidity vertically averaged over a range of pressure in the upper troposphere. The slope and intercept coefficients for equation (2) (a=31.50 and b=-0.1136 $K^{-1}$) are determined from a linear fit of the data in figure 4a and are consistent with the theoretically expected values (see SB93), further testifying to the validity of this simple relationship. In section 4 these coefficients will be used with equation (2) to interpret observed $T_{6.7}$ from TOVS in terms of an index of the vertically averaged upper tropospheric relative humidity.

---

1. In generating the TOVS archive, NESDIS applies a limb correction to convert the TOVS radiances viewed at a variety of zenith angles to equivalent nadir-view radiances ($\theta=0$). This procedure serves to remove the impact of varying viewing geometry upon the TOVS radiance field.

**b. Middle Tropospheric Water Vapor Radiances (7.3 μm)**

Next we evaluate the validity of the RSL model for interpreting middle tropospheric water vapor radiances. The sensitivity of the 7.3 μm brightness temperature ($T_{7.3}$) to variations in relative humidity for a typical tropical profile is shown in figure 1. Since the atmospheric attenuation is smaller at 7.3 μm than at 6.7 μm, the $T_{7.3}$ is sensitive to relative humidity averaged over a slightly lower layer of atmosphere (roughly 400 to 800 mbar).

Although this channel is located farther away from the center of the 6.3 μm band, the basic assumptions of the RSL model are still valid, namely that water vapor is the principal absorber and that the absorption can be modeled as a series of randomly-overlapped, strongly-absorbing, pressure-broadened lines. Hence equation (2) should also be valid for $T_{7.3}$, provided that the vertically averaged relative humidity $\bar{r}_m$ is now determined using weights for the middle tropospheric channel (listed in table 2). This is demonstrated in figure 4b which compares the synthetically calculated $T_{7.3}$ with the corresponding $\log(\bar{r}_m p_o/\cos\theta)$, employing the same set of input profiles used in figure 4a. The radiance calculations for the 7.3 μm water vapor channel exhibit good agreement with the RSL model (r=-0.95), demonstrating that equation (2) is also applicable for interpreting the $T_{7.3}$ in terms of a vertically averaged middle tropospheric relative humidity. The regression coefficients determined from a linear fit of the data are a=29.89 b=-0.0996 $K^{-1}$ and the scatter in $T_{7.3}$ for a given value of $\bar{r}_m$ is +/- 1.2 K, consistent with that obtained for the 6.7 μm channel.

For the July simulations shown above, virtually all the $T_{7.3}$ are at least 20 K colder than the surface temperature and therefore no surface correction is required. However, this is not the case year-round. During boreal winter, the atmosphere over the continental interiors of the northern mid-latitudes becomes relatively transparent due to the cold, dry conditions and correction for the surface contribution becomes necessary. An illustration of the surface correction procedure and its importance is provided in the next section for the lower tropospheric channel where its impact is moist significant.

**c. Lower Tropospheric Water Vapor Radiances (8.3 μm)**

Finally, we examine the application of the RSL model to the interpretation of TOVS lower tropospheric water vapor radiances. The 8.3 μm channel is located furthest from the center of the 6.3 μm band and therefore senses deepest into the atmosphere of the three water vapor channels. The sensitivity of $T_{8.3}$ to variations in relative humidity (figure 1) demonstrates that the $T_{8.3}$ are primarily influenced by the relative humidity below ~700 mb. Since the atmosphere is more transparent at 8.3 μm, the impact of the surface contribution is also greatest for this channel. The importance of this effect is demonstrated in figure 4c which compares the forward calculated $T_{8.3}$ (with no surface correction) versus log $(\bar{r}_L p_o/\cos\theta)$, where $\bar{r}_L$ is calculated using the lower tropospheric weighting curves listed in table 3. As suspected, a very poor correlation is evident (r=-0.53) indicating that the relationship predicted by the RSL model is not adequate for explaining the variations in the actual (i.e. unadjusted) $T_{8.3}$. Next, the simulated $T_{8.3}$ from figure 4c are adjusted to account for the surface contribution using the correction curve from figure 3 and the surface temperatures from the input profiles. The results are shown in figure 4d. Accounting for the surface contribution significantly improves the performance of the RSL model (r=-0.93). The relationship predicted by the RSL model now explains greater than 85% of the variation in brightness temperature and the scatter is reduced from > 7K to ~1.4 K. This result demonstrates two points: (i) that surface emission is an important component of the $T_{8.3}$ field; and (ii) when adjusted to account for the surface emission, the detailed calculation of $T_{8.3}$ agree well with the simple relationship predicted by the RSL model. The slightly greater scatter for the lower tropospheric channel (~1.4 K) compared to the middle and upper tropospheric channels (~1.2 K) may reflect greater vertical non-uniformity in relative humidity and lapse rate in the lower troposphere, particularly in the transition region from boundary layer to free atmosphere.

## 4. Climatologies of Upper, Middle, and Lower Tropospheric Humidity

This section presents near global maps of the annual-mean distribution, seasonal variation, and interannual variation of relative humidity in the upper, middle, and lower troposphere derived from the TOVS water vapor radiances. For a thorough description and error analysis of the TOVS water vapor radiance archive, see Wu et al. (1993), Kidwell (1991), and references therein. In this section, we use equation (2) in combination with the coefficients (a,b) derived for each channel to infer the distribution of upper, middle and lower tropospheric relative humidity from the TOVS 6.7, 7.3, and 8.3 μm $T_b$. For the middle and lower tropospheric channels, all $T_b$ which are not 20 K colder than the surface temperature are adjusted to correct for the surface contribution as described above. No adjustments are required for the upper tropospheric channel. Here the surface temperature is determined from monthly mean sea-surface temperature analyses (Reynolds, 1982) and hence results for the middle and lower tropospheric channel (which require surface temperature correction) are restricted to ocean surfaces.

### a. Geographic Variations

Figure 5a shows a map of the annual average upper tropospheric relative humidity (UTH) derived from the TOVS observed $T_{6.7}$ climatology for the period 1981-1991. In constructing this map, the UTH for each observation was determined from $T_{6.7}$ according to

$$UTH = \frac{\cos\theta}{p_o} e^{(31.50 - 0.1136\, T_{6.7})} \qquad (3)$$

where θ=0 for all locations and $p_o$ is determined from monthly-mean temperature climatologies. The UTH index was determined for each day and then averaged over the 11 year period. As demonstrated in section 3, the UTH is a measure of the relative humidity vertically averaged over a broad layer centered in the upper troposphere (roughly 200-500 mb). Figure 5b depicts the corresponding map of the annual average $T_{6.7}$. Recall that warmer $T_{6.7}$ correspond to a drier atmosphere, while colder $T_{6.7}$ correspond to a moister atmosphere. Both the UTH and $T_{6.7}$ maps show clearly the major features of the Hadley cell circulation with a moist (cold) intertropical convergence zone and a dry (warm) band extending over the subtropics. Moist regions are also observed along the mid-latitude storm tracks, over the western Pacific warm pool, and over monsoonal regions of South America and Africa where the relative humidity exceeds 50%. The driest regions are found off the west coasts of South America, Africa, and Australia where the annual mean relative humidity falls below 25%. Comparable spatial features are observed in radiosonde climatologies of relative humidity above 500 mbar (Piexoto and Oort, 1995), although the contrast between tropical and subtropical regions tends to be slightly reduced. The close resemblance between the $T_{6.7}$ and UTH fields emphasizes the fact that patterns of variation in UTH primarily reflect variations in $T_{6.7}$ and are not an artifact of the RSL model. In essence, the UTH is simply an exponential stretching to the brightness temperature field as predicted from the theoretical formulation of the RSL model.

Figure 5c shows a map of the annual average middle tropospheric humidity (MTH) index derived from the TOVS surface-corrected 7.3 µm brightness temperature climatology (denoted as $T^*_{7.3}$) for the same 11 year period. The MTH was computed using the coefficients determined in section 3,

$$MTH = \frac{\cos\theta}{p_o} e^{(29.89 - 0.0996\, T^*_{7.3})} \qquad (4)$$

The MTH corresponds to the relative humidity averaged over a layer extending from roughly 400-800 mb (see figure 1). The spatial patterns of the $T_{7.3}$ field (figure 5d) closely resemble those present in the MTH field. The basic features present in MTH are very similar to those observed for UTH. Most notable is the contrasting dry and moist regions over the tropics reflecting the upward and downward branches of the Hadley cell, and the moist areas associated with the mid-latitude storm tracks. On average, the MTH is slightly greater than the UTH suggesting that the relative humidity tends to decrease with height. The similarity between MTH and UTH indicates that dry (moist) regions in the middle troposphere are strongly correlated with dry (moist) regions in the upper troposphere. The close resemblance between the geographic patterns of relative humidity in the middle and upper troposphere, and the tendency for relative humidity to decrease with height are also noted in radiosonde climatologies (Piexoto and Oort, 1995).

Figures 5e and 5f depict maps of the annual average lower tropospheric relative humidity (LTH) and $T_{7.3}$. The LTH, which corresponds to the relative humidity averaged over a layer extending from roughly 700 mb to the surface, was determined from the TOVS surface-corrected 8.3 µm brightness temperature climatology ($T^*_{8.3}$)

$$LTH = \frac{\cos\theta}{p_o} e^{(25.35 - 0.0727\ T^*_{8.3})} \qquad (5)$$

Poleward of 45° latitude little information on LTH is available in the winter hemisphere because the $T_{8.3}$ are rarely 7 K cooler than $T_s$. Consequently the sampling of LTH for these latitudes is heavily biased towards summer conditions. In contrast to the upper and middle tropospheric channels, which show a strong similarity between the relative humidity and brightness temperature fields, the distribution of LTH differs noticeably from that of $T_{8.3}$. Indeed, although the LTH, MTH and UTH all show a broadly similar distribution, the $T_{8.3}$ exhibits a distinctly different spatial pattern from $T_{7.3}$ and $T_{6.7}$. This difference reflects the relatively strong surface contribution to the $T_{8.3}$, hence the spatial pattern of $T_{8.3}$ reflects a combination of the lower tropospheric humidity and surface temperature distributions. Although the primary features noted above are still present, the LTH tends to exhibit less spatial contrast than the relative humidity at higher levels, particularly in the tropics, suggesting that it is not as strongly influenced by large-scale atmospheric motions. Additionally, the relative humidity in the lower troposphere tends to be higher than that in the middle or upper troposphere, again implying that, when averaged over relatively thick layers, relative humidity tends to decrease with height in a fairly uniform manner.

**b. Seasonal Variations**

Figure 6 shows the seasonal variation in UTH (top), MTH (middle), and LTH (bottom) averaged over the period 1981-1991. The seasonal variations in relative humidity closely resemble each other in the middle and upper troposphere and emulate the seasonal patterns of variation in the Hadley circulation. Similar, though slightly weaker and less coherent variations are observed in the lower troposphere. The largest changes occur over the tropics where distinct shifts following the seasonal progression of the ITCZ are clearly evident. Weaker seasonal variations are noted over the extratropics. For the upper and middle troposphere, maximum relative humidity occurs along the ITCZ from June-September and is associated with a corresponding minimum over the southern hemisphere subtropics during the same period; i.e. the wettest period in the tropics coincides with the driest period in the subtropics. A secondary pair of maxima/minima occurs during the southern hemisphere winter, November-February, with the largest relative humidities along the equator and the lowest relative humidities over the northern hemisphere subtropics. Both pairs of minima/maxima typically lag the solstice period by 1.5-2 months. Meridional gradients in relative humidity are smallest during the equinox periods, reflecting a weaker zonal-mean Hadley circulation. Also note that the southern hemisphere subtropical dry zone is more expansive meridionally and exibits less seasonal shift than does its counterpart in the northern hemisphere. Radiosonde climatologies display similar seasonal variations in relative humidity (Piexoto and Oort, 1995).

To illustrate the zonal asymmetries in the seasonal variation, figure 7 shows a map of the annual cycle in UTH. A fourier analysis was performed to isolate the annual harmonic and determine its amplitude and phase. The amplitude is given by the length of the arrow in relation to the scale on the lower right. The phase corresponds to the time of maximum UTH and can be determined from the orientation of the arrows with respect to a clock. Arrows pointing toward 1 o'clock correspond to a maximum in January, those toward 2 o'clock correspond to a maximum in February, etc. The greatest seasonal variations (> 10%) are observed over tropical monsoonal regions such as Central America, Brazil, India, and central Africa and exhibit a clear half-year phase difference between regions north and south of the equator. Large seasonal variations also occur over the southeast and northwest quadrants of the tropical Pacific with a similar half-year phase lag. Over the northern extratropics, a distinct land-ocean asymmetry is present. The UTH tends to be greatest during winter over North American and Asian continents, while over the Pacific and Atlantic oceans maximum UTH is observed during summer. Over the southern extratropics, little seasonal variablity is observed.

**c. Interannual Variations**

To examine interannual variations in relative humidity we performed empirical orthogonal function (EOF) analysis on monthly UTH anomalies within the 30N-30S tropical belt. Figure 8 shows the spatial pattern of the leading EOF of the interannual variation of UTH (top) and the corresponding expansion coefficient (bottom) for the period 1981-1991. For brevity only results for UTH are shown as the MTH and LTH patterns exhibit similar features. A similar analysis of the interannual variation in $T_{6.7}$ was recently described by Bates and Wu [1995]. A notable difference between this study and that of Bates and Wu, is that the latter used an adjusted $T_{6.7}$ anomaly time series in an attempt to remove intersatellite calibration differences. Although no such adjustment was performed here, the resulting expansion coefficient and EOF pattern are in good agreement with the corresponding quantities determined by Bates and Wu. Thus for our purpose, the occurence of slight intersatellite calibration differences does not appear to significantly affect the interpretation.

The first EOF, which explains 11% of the total variance, depicts a pattern of spatial variation commonly associated with the El Nino/ Southern Oscillation (ENSO). This is supported by the expansion coefficients which peak in the early part of 1983 and 1987, and late 1991 all of which correspond to recognized ENSO warm phases. A coherent pattern of UTH anomalies extends throughout the tropical belt. Maximum positive anomalies are located over the central tropical Pacific where a wetter upper troposphere is associated with the eastward shift of convection. Maximum negative anomalies are centered over Indonesia, reflecting a decrease in convective activity over this region during ENSO events and the associated reduction in the Walker circulation. The double lobe apparent in the positive anomaly along the equatorial Pacific is believed to be associated with the late stages of the 1983 ENSO (Bates and Wu, 1995). To the north and south of this moist band are negative (dry) anomalies, the strongest of which occurs over the northern subtropics, suggesting an increase in atmospheric subsidence in these regions. An exception to this is the positive anomaly in the southern subtropics off the west coast of South America. An interesting feature of these patterns is that outside the central Pacific the polarity of the anomaly pattern reverses. For example, from $60^o$ W to $120^o$ E negative (dry) anomalies occur over much of the equatorial belt ($15^o$ N to $15^o$ S) while positive (moist) anomalies dominate the subtropics ($15^o$ - $30^o$ N and S). This suggests corresponding changes in the patterns of tropical circulation at these scales during ENSO events.

## 5. Sensitivity Analysis

This section examines the sensitivity of the relative humidity indices presented in the previous section to the profile data set used to tune the RSL model. The coefficients in equation (3) were determined by comparing synthetic radiances calculated from representative profiles of temperature and moisture with the corresponding vertically averaged relative humidity from the same profiles. This procedure essentially represents a tuning of the RSL relationship to the CIMSS transmittance calculations, using ECMWF analyses to provide the input temperature and moisture profiles. Here we demonstrate that the UTH index derived from the RSL model is insensitive to the particular profile data set used in the tuning.

To examine the sensitivity of RSL model to this tuning procedure, we compare the UTH determined using the coefficients from section 3 (hereafter referred to as the "reference" UTH) with that determined using two new pairs of coefficients, each derived from different profile data

sets - National Meteorological Center analyses (NMC), and TOVS Initial Guess Retrieval radiosonde profiles (TIGR). The NMC data set represents a collection of roughly 8,000 profiles taken from the operational analysis produced for Sept. 15, 1990, using all profiles equatorward of $60^o$ latitude. The TIGR profile data set is collection of roughly 1,700 soundings carefully selected from a much larger data set (~100,000) with the objective that they should be representative of the global atmosphere at any time and place (Monine et al., 1987). Table 4 lists the coefficients obtained from both new profile data sets. Figure 9 compares the annual mean climatology of the reference UTH with that obtained from both pairs of new coefficients. Each data point in these figures corresponds to the annual mean UTH from a $2.5^o$ grid box between 60N and 60S. Statistics from this comparison are listed in table 4. There are two points to be made with this analysis. One is that spatial variations in the relative humidity indices are not dependent upon the particular set of coefficients used, but rather are a reflection of variations in the water vapor radiance field. This is evidenced by the near-perfect correlation ($r > 0.999$) between the reference UTH and that determined using either of the new coefficients. The second point is that, although systematic biases result when different coefficients are used, these biases are small - typically less than 5%. The rms differences are also on the order of 5%. Thus, although the precise value of the humidity inferred from the TOVS radiances depends slightly upon the profile data set used to tune the RSL model, the uncertainty is small relative to the magnitude of the regional and temporal variations of relative humidity. This insensitivity reflects both a certain degree of consistency between the various profile data sets as well as the fundamental nature of the physical principles upon which the theoretical model is based. A similar degree of insensitivity is apparent for the middle and lower tropospheric humidity indices (not shown). In essence, the RSL model provides a simplified relation derived from radiative transfer theory, tuned to agree with the CIMSS transmittance calculations, and is relatively insensitive to the particular profile data set used in the tuning.

## 6. Discussion

Satellite observations of the upwelling radiance in the 6.3 μm water vapor absorption band are used to describe the geographic distribution and temporal variation of relative humidity. The combination of daily, near-global coverage with a coherent archive dating back to 1979 makes TOVS water vapor radiances a valuable source of information on tropospheric water vapor. The water vapor channels, located at 6.7 μm, 7.3 μm, 8.3 μm, are sensitive to the amount of relative humidity integrated over broad layers centered in the upper (200-500 mb), middle (400-800 mb), and lower troposphere (700mb - surface), respectively. To facilitate the interpretation of the TOVS radiances in terms of a more familiar water vapor quantity, a random strong line model is applied. This model, based upon an assemblage of randomly overlapped, strongly absorbing, pressure broadened lines, indicates that, accurate to about 1 K, the brightness temperature measures essentially the logarithm of the mean relative humidity averaged over broad layers of the troposphere subject to a straight-forward allowance for viewing angle, a minor correction for the mean temperature structure, and an adjustment for the surface contribution, if necessary. The validity of this simple relationship is supported by comparison with detailed radiative transfer calculations using the CIMSS transmittance model and known profiles of temperature and moisture.

The climatological distribution of relative humidity is similar in all three layers, reflecting well-known patterns of the large-scale atmospheric circulation. Peak relative humidities occur along the ITCZ, with minimum values over the subtropics, and a second band of maxima over the mid-latitude storm tracks. On average, the UTH index is smaller than the MTH and LTH indices, suggesting a general tendency for the relative humidity to decrease with height. Seasonal variations in UTH and MTH are greatest over tropics where they emulate the migration of the Hadley circulation. Weaker seasonal variations are observed over extratropics. Interannual variations demonstrate a clear ENSO signal, with coherent pattern of anomalies extending throughout the tropics which reflect changes in the Hadley and Walker circulation patterns.

The TOVS water vapor archive provides valuable information on tropospheric water vapor suitable for comparing with GCMs. Although the water vapor radiances are strongly correlated to the layer-mean relative humidities, the most appropriate and consistent method for using satellite observations to evaluate GCMs is to follow a "profile to satellite approach" (e.g. Soden and Bretherton, 1994; Schmetz and van de Berg, 1994; Salathe et al., 1994 Chen et al., 1995). That is, model profiles of temperature and moisture should be inserted into a radiative transfer model to simulate the radiance which would be observed by the satellite under those conditions. Then both the observed and model-simulated radiances can be interpreted in terms of relative humidity using the RSL model as outlined in this paper.

Table 1: Weights for estimating $\bar{r}_u$ from r at standard pressure levels.

| Standard level (mb) | $p_{6.7}=250$ | $p_{6.7}=300$ | $p_{6.7}=350$ | $p_{6.7}=400$ |
|---|---|---|---|---|
| 150 | 0.18 | 0.08 | 0.04 | 0.02 |
| 200 | 0.22 | 0.15 | 0.09 | 0.05 |
| 250 | 0.24 | 0.20 | 0.14 | 0.10 |
| 300 | 0.23 | 0.27 | 0.25 | 0.20 |
| 400 | 0.11 | 0.22 | 0.29 | 0.29 |
| 500 | 0.02 | 0.07 | 0.16 | 0.25 |
| 700 | 0 | 0.01 | 0.03 | 0.09 |

Table 2: Weights for estimating $\bar{r}_m$ from r at standard pressure levels.

| Standard level (mb) | $p_{7.3}=350$ | $p_{7.3}=400$ | $p_{7.3}=500$ | $p_{7.3}=600$ |
|---|---|---|---|---|
| 200 | 0.14 | 0.09 | 0.04 | 0.01 |
| 250 | 0.17 | 0.11 | 0.07 | 0.03 |
| 300 | 0.25 | 0.21 | 0.12 | 0.07 |
| 400 | 0.26 | 0.25 | 0.19 | 0.12 |
| 500 | 0.15 | 0.22 | 0.26 | 0.22 |
| 700 | 0.03 | 0.10 | 0.22 | 0.27 |
| 850 | 0 | 0.02 | 0.08 | 0.22 |
| 1000 | 0 | 0 | 0.02 | 0.06 |

Table 3: Weights for estimating $\bar{r}_l$ from r at standard pressure levels

| Standard level (mb) | $p_{8.3}=700$ | $p_{8.3}=800$ | $p_{8.3}=900$ | $p_{8.3}=950$ |
|---|---|---|---|---|
| 300 | 0.03 | 0 | 0 | 0 |
| 400 | 0.07 | 0.07 | 0.05 | 0.03 |
| 500 | 0.23 | 0.20 | 0.16 | 0.12 |
| 700 | 0.35 | 0.36 | 0.31 | 0.27 |
| 850 | 0.26 | 0.29 | 0.35 | 0.40 |
| 1000 | 0.06 | 0.08 | 0.13 | 0.18 |

Table 4: Sensitivity of UTH to regression coefficients.

| Profile set | a | b | correlation | rms | bias |
|---|---|---|---|---|---|
| TIGR | 30.78 | -0.1113 | 0.9999 | 3.6% | -3.2% |
| NMC | 28.70 | -0.1030 | 0.9991 | 2.24% | -1.05% |
| Reference | 31.50 | -0.1136 | 1.0 | 0.0 | 0.0 |

The correlation, rms difference, and bias are calculated with respect to the reference UTH (see text for details).

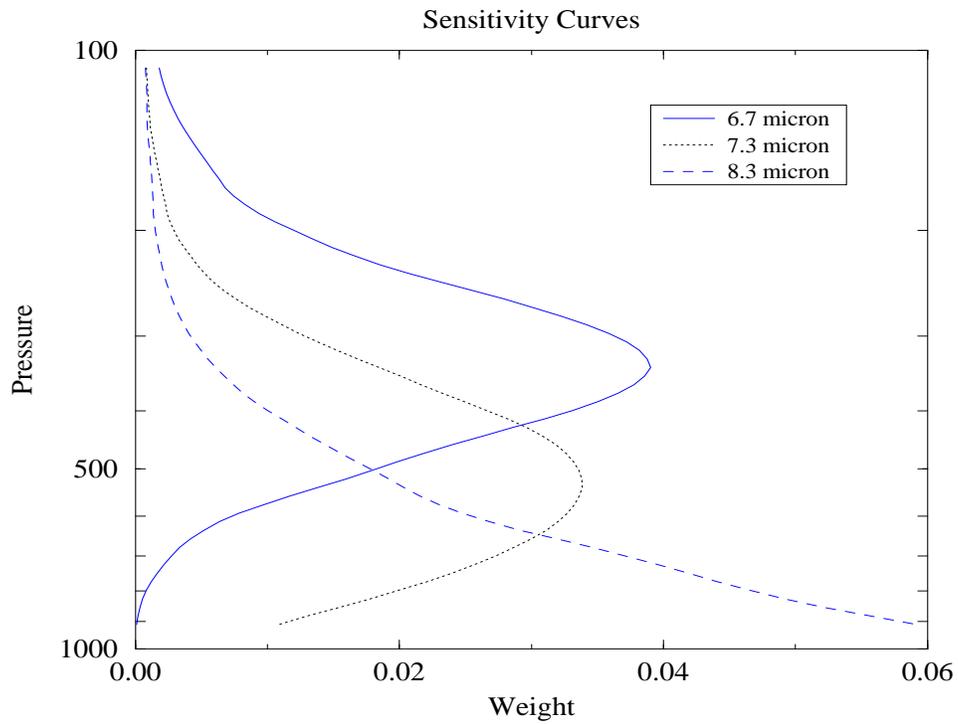

**Figure 1** Sensitivity of $T_{6.7}$ (solid), $T_{7.3}$ (dotted), and $T_{8.3}$ (dashed) to local variations in relative humidity. The curves are normalized such that the sum of weights over pressure is equal to unity.